# Studies on Neon irradiated amorphous carbon using X-ray Diffraction technique

A.Sarkar[*], K. Dasgupta[#], P. Barat[*,@], P. Mukherjee[*], D. Sathiyamoorthy[#]

[*]*Variable Energy Cyclotron Centre, 1/AF Bidhan Nagar, Kolkata– 700 064, India*

[#]*Materials Group, Bhabha Atomic Research Centre, Mumbai– 400 085, India*

**Abstract**

Two sets of amorphous carbon materials prepared at different routes are irradiated with swift (145 MeV) heavy ion ($Ne^{6+}$). The structural parameters like the size of ordered grains along c and a axis i.e. $L_c$ & $L_a$, the average spacing of the crystallographic planes (002) i.e. $d_{002}$ and the fraction of the amorphous phase of the unirradiated and the irradiated samples are estimated by X-ray diffraction technique. The fraction of the amorphous phase is generally found to increase with the irradiation dose for both sets of the samples. The estimated $d_{002}$ and $L_a$ values are found to be almost unaffected by irradiation. The estimated values of $L_c$ corroborate with the increase of disorder in both sets of the samples with the increasing dose of irradiation.

**Keywords:** X-ray Diffraction, Amorphous Carbon, Irradiation

**Introduction**

The effects of particle irradiation in solids have been a field of intense research since 1950s. The early studies focused on radiation damage in materials of nuclear technology to understand the variation of their physical and mechanical properties due to irradiation. The allotropes of carbon have been the subject of many early studies; in particular graphite has attracted much attention as a nuclear reactor material in fission

---

[@] Corresponding author, Email: pbarat@veccal.ernet.in

reactors. Moreover in recent years the development of proper structural materials for fusion reactors where carbon is considered to be one of the potential candidate material, is a challenging job to the nuclear technologists. From the renewed interest in radiation damage in carbon structures a need for the generation of data about the behavior of carbon materials under particle irradiation has arisen.

The use of graphite as moderator in high temperature nuclear reactors is well known [1]. However, in low temperature thermal nuclear reactors, graphite cannot be used as a moderator due to the possibility of sudden release of the stored energy (Wigner energy) [1, 2] in it, resulting from the displacement of atoms from its lattice position during irradiation. Calculation shows that this stored energy may get released suddenly around 200°C and can lead to an accident in the thermal nuclear reactors [1,2]. Use of amorphous carbon as a moderator can eliminate the Wigner energy effect. To develop such a novel carbonaceous material to be used in the nuclear reactor, two sets of samples of amorphous carbon have been prepared following different routes. It is interesting to study whether the particle irradiation causes formation of any crystalline phase in the amorphous carbon samples. This study will help us to be safe in using the amorphous carbon sample as moderator in the low temperature thermal nuclear reactors. During irradiation the incident particles deposit kinetic energy in the localized spatial region of the irradiated samples. There is a possibility that this deposited energy can cause localized ordered arrangement in the disordered structure of the atoms in the amorphous samples leading to crystallinity.

In the present work, we have carried out irradiation with 145 MeV $Ne^{6+}$ ions on the amorphous carbon samples at different doses. We have characterized the structural parameters like the size of the ordered grains along c and a axis i.e. $L_c$ & $L_a$, the average spacing of the crystallographic planes (002) i.e. $d_{002}$ and the fraction of amorphous phase of the unirradiated and the irradiated samples as a function of dose using the X-ray diffraction (XRD) technique. The damage profile as a function of depth from the surface of these samples has been characterized in terms of displacement per atom (dpa) for different doses.

**Experimental**

*Sample preparation:*

Two types of amorphous carbon have been prepared. Sample A was prepared with 15% carbon black dispersed in phenolic resin. The sample was carbonized at $1500^0$C with a heating rate of $100^0$C /hr in argon atmosphere. Sample B was prepared from chopped PAN fiber in resin matrix. The sample was then carbonized at $1000^0$C in argon atmosphere with a heating rate of $60^0$C/hr.

*Irradiation experiment:*

The samples A and B were mounted on an aluminum flange and then irradiated with 145 MeV $Ne^{6+}$ ions from Variable Energy Cyclotron (VEC), Kolkata, India. The irradiation doses were $1.0\times10^{13}$, $5.0\times10^{13}$, $1.5\times10^{14}$ $Ne^{6+}/cm^2$. The flange used in irradiation was cooled by continuous flow of water. During irradiation, the temperature of the samples did not rise above 313K as monitored by a thermocouple placed inside the groove of the flange in close proximity of the sample. The dpa was obtained by Monte-Carlo simulation technique using the code SRIM 2000 [3].

*X-ray diffraction:*

A Philips PW1710 X-ray diffractometer was used to record X-ray intensities scattered from the examined samples. Copper $k_\alpha$ (wavelength $\lambda=1.5418$ Å) radiation (40kV, 40 mA) was used as an X-ray source. Samples were scanned in a step-scan mode ($0.02°$/step) over the angular range ($2\theta$) of $15°$ to $70°$. X-ray diffraction data were collected for 2.5 sec at each step.

*Resistivity measurement:*

Resistivity at room temperature of the samples were measured using four probe technique with a HP 34220A nanovoltmeter with resolution of 0.1 nV and a Keithley 224 programmable current source. 1 mA current was employed.

**Method of analysis:**

Studies of amorphous materials represent a large and important emerging area of materials science. It is an area which is not amenable to the most of the conventional theoretical techniques of solid-state physics as there is no periodicity to simplify the mathematics. Due to the lack of periodicity, extraction of structural information from amorphous materials becomes very difficult. In this work we have used XRD technique to analyze the variations of the structural parameters of amorphous carbon due to irradiation. XRD is a very useful and simple technique to understand the structural details of the solid-state substances. Incident X-ray interacts with large volume of the material at a time and an average property of the material can be characterized rather than the local property. This makes XRD a powerful technique for studying the disordered materials which is inherently heterogeneous and where the estimation of average property has got practical significance.

Fig. 1 shows a typical XRD profile for the unirradiated sample A. The broad peaks and the diffused nature of the XRD profile indicate the presence of disorder in the sample [4]. The first peak of the profile corresponds to the (002) peak of the ordered hexagonal graphite structure. While the second peak corresponds to the (100) and (101) peaks, which is often called the (10) band. Before carrying out any analysis the diffraction profiles were corrected for polarization and absorption using the techniques described in the literature [4,5]. Fig. 2 shows a typical normalized corrected intensity curve [Curve A] for the unirradiated sample A plotted against s $\left(s = \frac{2\sin\theta}{\lambda}\right)$ along with the coherent scattering [Curve B], the incoherent Compton scattering [Curve C] and the total independent scattering i.e. sum of the curves B and C [Curve D]. Since the observed intensity curve in Fig. 1 can be affected by many random factors, such as sample packing, it can not be used directly either for quantitative analysis or for comparison of spectra of the samples irradiated with different doses. To eliminate these problems, the intensity was converted to electronic units by normalizing the profile with a suitable factor so that at large $s$ the normalized curve [Curve A] oscillates about the Curve D [4,6,7]. The reduced intensities [4,6] of the samples: $I = i(s) = \frac{A-C}{B}$ were then calculated from the normalized corrected intensities. Fig. 3 shows the typical reduced intensity plot against s for the unirradiated sample A. Further quantitative analysis is based on this reduced intensity profile rather than the observed one.

*Fraction of amorphous carbon:*

The (002) peak of the sample in the observed diffraction profile (Fig. 1) is almost at the same position as that observed in crystalline layered hexagonal graphite structure [PDF# 751621]. However, it is diffused and broadened indicating that the sample exists in a form which is intermediate between the amorphous and the crystalline states [4,8]. These types of structures are built up from the individual graphite layers arranged parallel to one another slightly greater the normal graphite spacing (crystalline) but random in translation parallel to the layer and rotation about the normal (amorphous) [8] and known as random layer lattice structure.

Hence, the diffraction pattern of such a material consists of two types of reflections – crystalline type reflections and diffuse two dimensional lattice reflections (amorphous contribution). Therefore the reduced intensity ($I$) can be expressed as the sum of two separate contributions from the crystalline carbon ($I_{Cr}$) and the amorphous carbon ($I_{am}$),

$$I = I_{Cr} + I_{am}. \qquad (1)$$

The fraction of amorphous carbon does not contribute to the peak intensity and is only reflected in the background of the intensity pattern. According to Franklin [5] the intensity contributed by the amorphous carbon is constant over the whole scattering range and equal to the fraction of the amorphous carbon ($x_A$). Thus, the reduced intensity can be expressed as

$$I = I_{Cr} + x_A. \qquad (2)$$

For the first peak of the diffraction pattern *i.e.* for (002) reflection, the above intensity equation can be written as

$$I = I_{002} + x_A \qquad (3)$$

Warren theoretically calculated the intensity equations for random layer lattices in terms of the layer dimension and the position of the related crystalline reflection [8]. Applying

Warren results in case of carbon Franklin [5] found out the expression for the reduced intensity of the (002) reflection as

$$I_{002} = \left(\frac{0.0606}{s^2}\right) \times \sum_n \left(p_n \times \frac{\sin^2(\pi n d_n s)}{(n \sin^2(\pi d_n s))}\right) \quad (4)$$

Where $d_n$ is the inter-layer spacing on the groups of N parallel layers, and $p_n$ is the fraction of the total carbon contained in such groups. It is obvious from the intensity equation that the fraction of the carbon, which does not take part in the layer structure, does not contribute to the (002) band. According to our assumption the fraction of crystalline carbon (layered structure) present in the sample is $(1 - x_A)$. Thus the actual reduced intensity due to (002) reflection reduces to

$$I_{002} = (1 - x_A) \times \left(\frac{0.0606}{s^2}\right) \times \sum_n \left(p_n \times \frac{\sin^2(\pi n d_n s)}{(n \sin^2(\pi d_n s))}\right) \quad (5)$$

Using this in Eq. (3) we arrive at

$$\frac{I - x_A}{1 - x_A} \times \frac{s^2}{0.0606} = \sum_n \left(p_n \times \frac{\sin^2(\pi n d_n s)}{(n \sin^2(\pi d_n s))}\right) \quad (6)$$

The right-hand side of the above equation is a periodic polynomial expression with the maximum occurring at $d_n s^i_{max} = i$ (any integer). Its first maximum occurs at $d_n s^i_{max} = 1$. This peak corresponds to the (002) band in the reduced intensity curve. Since each item in this polynomial expression, consequently the summation is symmetrical around the maximum $s^i_{max}$, the left–hand side of the equation should have the same symmetrical profile. This fact has been exploited to determine the fraction of amorphous carbon ($x_A$) in the samples [5]. The expression in the left-hand side of the Eq. (6), $(I - x_A)/(1 - x_A) \times (s^2/0.0606)$, is very sensitive to $x_A$.

To determine the fraction of amorphous carbon in the samples the expression $(I-x_A)/(1-x_A) \times (s^2/0.0606)$ is calculated from the reduced intensity ($I$) and plotted as a function of $s$. $x_A$ is varied to obtain the best symmetrical plot of $(I-x_A)/(1-x_A) \times (s^2/0.0606)$ vs. $s$ by fitting a Loretzian curve. Fig. 4 shows a typical most symmetric plot of $(I-x_A)/(1-x_A) \times (s^2/0.0606)$ vs. $s$ for the unirradiated sample A for $x_A$=0.76. This value of $x_A$ corresponds to the fraction of amorphous carbon in the sample [5]. The $s$ value corresponding to the maximum of the plot ($s_{max}^i$) is obtained from the fit. Assuming that the inter-layer spacing is the same for all size groups and equal to the mean inter-layer spacing of the crystalline structure ($d_{002}$) we get the lattice spacing $d_{002}$ from the relation $d_{002} s_{max}^i = 1$.

*Average stacking height and crystalline diameter:*

XRD is the most common analytical technique used for determining the structure of ordered and disordered carbons [5,9-12]. Ordered graphite has the hexagonal structure (space group P6$_3$/mmc) with carbon layers having the ABAB-stacking along the c-axis, although rhombohedral 3R graphite with ABCABC-stacking is also possible. Disorder can occur due to variety of reasons, including the presence of local 3R stacking, random shifts between adjacent layers, unorganized carbons which are not part of layer structure, and strain in the layers. These disorders affect the ordered grain size $L_c$ and $L_a$ along c-axis and a-axis respectively. $L_c$ and $L_a$ can be approximately calculated from the (002) and (10) band respectively using an empirical expression first derived by Scherrer [13,14], which is

$$L = \frac{K\lambda}{B\cos\theta} \tag{7}$$

Where B and $\theta$ corresponds to the full width at half maximum (FWHM) and the position of the peak respectively. K is a constant depending on the reflection plane and equals to 0.89 and 1.84 for (002) band (10) band [15-18] respectively.

**Results and Discussions**

Many studies have been made on the effect of heavy ion irradiation in crystalline materials. On the contrary very few literatures can be found on the irradiation effect in amorphous materials. In this work we have extensively studied the irradiation effect caused by swift heavy ion on two sets of amorphous carbon samples. The radiation damage in the samples is assayed by the damage energy deposition causing displacements of atoms. The $d_{002}$, $L_c$ and $L_a$ values obtained for the unirradiated and the irradiated samples are listed in Table 1. The estimated size of ordered grains in both the samples is found to decrease along c-axis, but remains almost unaffected along the a-axis as a function of dose. The $d_{002}$ values also did not change much due to irradiation. The constancy of $d_{002}$ values and the decrease of $L_c$ with the dose of irradiation suggest that the spacing of the lamellar planes of hexagonal structure of graphite did not change due to irradiation but their orientations with respect to each other got randomized. $L_a$ and $L_c$ measure the long range order of these lamellar structures along a and c axis respectively. The constancy of $L_a$ and the reduction of $L_c$ with the irradiation can be attributed to the fact that the presence of the strong covalent ($\sigma$ and $\pi$) bonding along the a-axis and the weak van der Waals bonding along the c-axis. The change in the fraction of amorphous carbon in the samples due to increasing dose of irradiation is shown in Fig. 5. It is seen that the fraction of amorphous carbon in sample A gradually increased due to irradiation. In case of sample B, the fraction of amorphous carbon has increased substantially at initial doses but there is some increase of ordering at the highest dose of irradiation. Fig. 6a and 6b shows the variation of $L_c$ and $L_a$ with the fraction of amorphous carbon for the samples A and B respectively. It is seen that for both the samples $L_c$ and $L_a$ vary linearly with the fraction of amorphous carbon. The fraction of amorphous phase present in the samples is a measure of the degree of disorder in it. The linear dependence of $L_c$ and $L_a$ clearly demonstrates that $L_c$ and $L_a$ can also be a measure of the disorder.

The range of 145 MeV $Ne^{6+}$ ion in the samples is 143 μm. The kinetic energy of the incident projectile ($Ne^{6+}$) is primarily deposited on the target by ionization and nuclear energy loss. The measured electrical resistivities of the samples were of the order of $10^{-4}$ Ωm. This is quite appreciable to quench the ionization caused by irradiation almost instantaneously. Hence, it may be argued that the observed increase of disorder in the sample due to irradiation did not cause due to ionization but due to nuclear energy loss.

Neon being heavy ion, transfers sufficient kinetic energy to the primary knock on atoms which in turn produce displacement cascades. As the primary knock on proceeds through the sample, loosing energy in successive collisions, the displacement cross-section increases [19]. This displacement of atoms increased the disorder in the samples. For the highest dose ($1.5 \times 10^{14}$ $Ne^{6+}/cm^2$) of irradiation the damage is found to be $0.3 \times 10^{-3}$ dpa, as calculated from SRIM 2000 [3]. The sample B is more crystalline in the unirradiated state. During irradiation, sufficient amount of energy is transferred in the system for electronic excitation and formation of defects. During deexcitation of electrons, enormous amount of energy dissipation (roughly 100-1000 times higher than the stored energy) leads to disordering upto a certain dose. But, at the highest dose, the export of entropy is so high that it may exceed the internal entropy production and leads to reduction of the total entropy of the system [20]. This can result in ordering in the sense of statistical thermodynamics and the fraction of amorphousity comes down. This ordering effect under irradiation at high doses is explained as self-organized phenomenon in the literature [21, 22].

Moreover, it is seen that the behavior of the carbonaceous samples under irradiation is dependent on the nature of the precursor used for preparing these sample. The process and precursor adopted for making sample A, with high fraction of amorphous carbon, is more suitable for the use as a moderator in thermal nuclear reactors as it is seen that the fraction of amorphous carbon has increased with irradiation reducing the possibility of Wigner energy storage.

**Conclusions**

Two sets of novel amorphous carbon materials are irradiated with high-energy $Ne^{6+}$ ions and the structural parameters of the unirradiated and irradiated samples are estimated by XRD profile analysis. The fraction of amorphous carbon is generally found to increase with the irradiation dose for both the samples. The estimated values of the $d_{002}$ and $L_a$ did not change much due to irradiation. The estimated values of $L_c$ corroborate with the increase of disorder in both the samples with the increasing dose of irradiation.

Table 1. Variation of $d_{002}$, $L_c$ and $L_a$ with the irradiation dose.

| Dose (Ne$^{6+}$/cm$^2$) | Sample A | | | Sample B | | |
|---|---|---|---|---|---|---|
| | $d_{002}$ (Å) | $L_c$ (Å) | $L_a$ (Å) | $d_{002}$ (Å) | $L_c$ (Å) | $L_a$ (Å) |
| Unirradiated | 3.50 | 11.7(4) | 7.7(4) | 3.48 | 12.0(3) | 8.2(3) |
| 1.0×10$^{13}$ | 3.53 | 10.2(2) | 7.3(2) | 3.50 | 10.7(2) | 7.8(4) |
| 5.0×10$^{13}$ | 3.52 | 10.5(2) | 7.5(3) | 3.55 | 10.0(3) | 7.5(3) |
| 1.5×10$^{14}$ | 3.56 | 9.6(3) | 7.0(3) | 3.52 | 10.3(2) | 7.7(4) |

**Figure Captions:**

Fig. 1. XRD profile of the unirradiated sample A.

Fig. 2. Normalized corrected intensity curve [Curve A] for the unirradiated sample A along with the variation of the coherent scattering [Curve B], the Compton scattering [Curve C] and the total independent scattering [Curve D] against s.

Fig. 3. Reduced intensity curve for the unirradiated sample A

Fig. 4. Determination of fraction of amorphous carbon ($x_A$) in the unirradiated Sample A.

Fig. 5 Variation of fraction of amorphous carbon with the irradiation dose.

Fig. 6 Variation of $L_c$ and $L_a$ with the fraction of amorphous carbon.

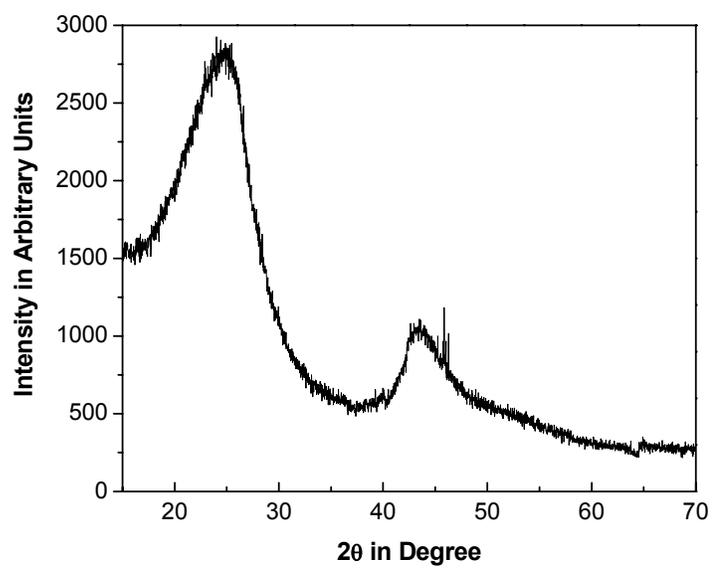

Fig. 1. XRD profile of the unirradiated sample A.

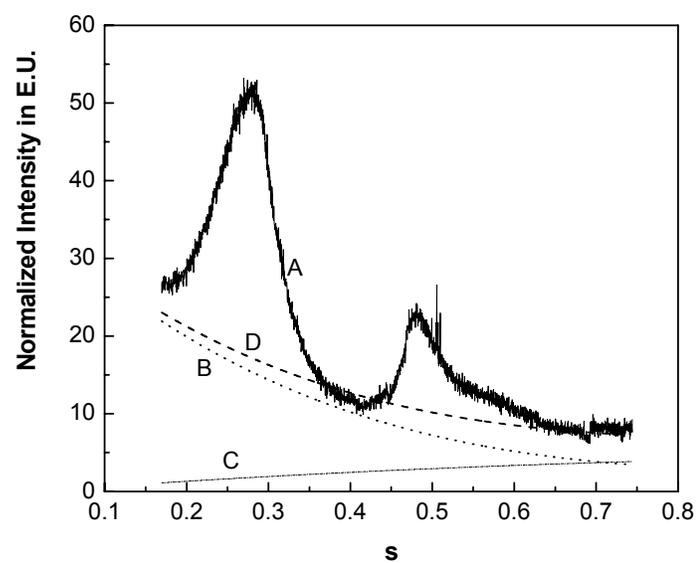

Fig. 2. Normalized corrected intensity curve [Curve A] for the unirradiated sample A along with the variation of the coherent scattering [Curve B], the Compton scattering [Curve C] and the total independent scattering [Curve D] against s.

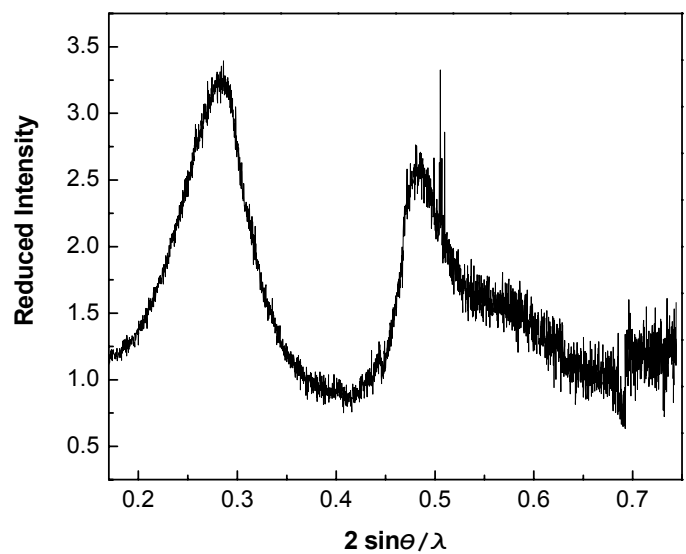

Fig. 3. Reduced intensity curve for the unirradiated sample A

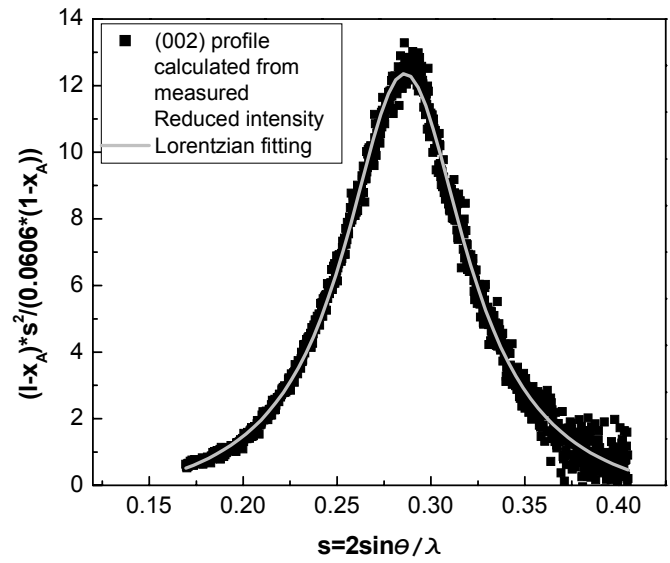

Fig. 4. Determination of fraction of amorphous carbon ($x_A$) in the unirradiated Sample A.

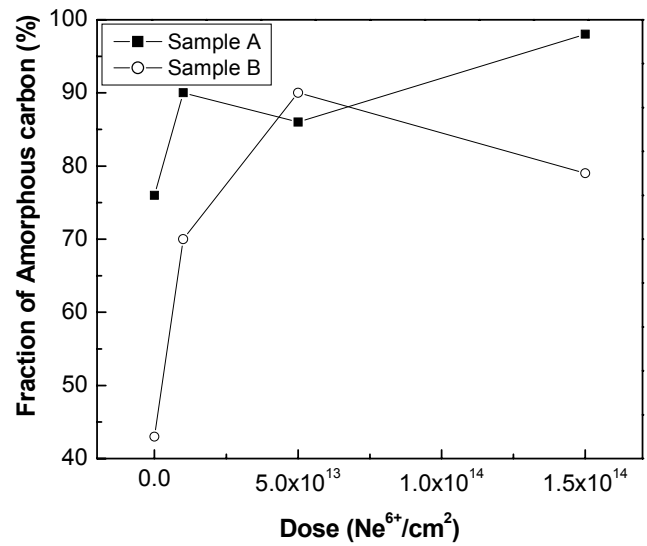

Fig. 5 Variation of fraction of amorphous carbon with the irradiation dose.

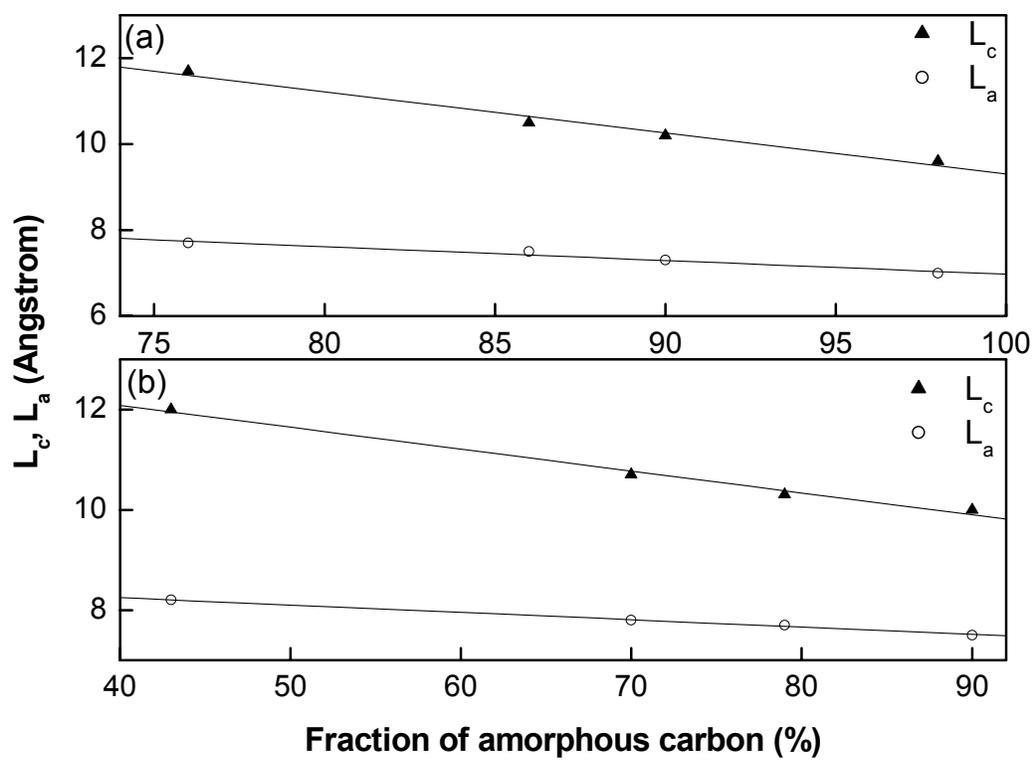

Fig. 6 Variation of $L_c$ and $L_a$ with the fraction of amorphous carbon.